\documentclass[pra,reprint,superscriptaddress,longbibliography,nofootinbib]{revtex4-1}

\usepackage{amsmath,amssymb}
\usepackage[euler]{textgreek}

\usepackage{psfrag}
\usepackage{graphicx}
\usepackage{color}
\usepackage[usenames,dvipsnames]{xcolor}
\usepackage{pifont}
\usepackage{fourier}
\usepackage{xr}
\usepackage{stmaryrd}
\usepackage{cancel}

\usepackage{fancyhdr}

\usepackage[process=auto]{pstool}

\usepackage[normalem]{ulem}
\usepackage{euscript,array}

\newcommand{\bea}{\begin{eqnarray}}
\newcommand{\eea}{\end{eqnarray}}
\newcommand{\beq}{\begin{equation}}
\newcommand{\eeq}{\end{equation}}

\definecolor{redg}{rgb}{1,0,0}
\definecolor{blueg}{rgb}{0.22,0.33,0.64}
\definecolor{greeng}{rgb}{0,0.63,0.29}
\definecolor{orangeg}{rgb}{0.96,0.47,0.13}

\DeclareMathAlphabet\mathbfcal{OMS}{cmsy}{b}{n}

\newcommand{\hlambda}{\hat \lambda}

\begin{document}

\title{Ground state modulations in the ${\mathbb C}P^{N-1}$ model}

\author{Antonino Flachi}
\affiliation{Department of Physics \& Research and Education Center for Natural Sciences, Keio University, 4-1-1 Hiyoshi, Kanagawa 223-8521, Japan}
\author{Guglielmo Fucci} 
\affiliation{Department of Mathematics, East Carolina University, Greenville, NC 27858, USA}
\author{Muneto Nitta}
\affiliation{Department of Physics \& Research and Education Center for Natural Sciences, Keio University, 4-1-1 Hiyoshi, Kanagawa 223-8521, Japan}
\author{Satoshi Takada} \affiliation{Institute of Engineering, Tokyo University of Agriculture and Technology, 2-24-16, Naka-cho, Koganei, Tokyo 184-8588, Japan}
\author{Ryosuke Yoshii}\affiliation{Department of Physics, Chuo University, 1-13-27 Kasuga, Bunkyo-ku, Tokyo 112-8551, Japan}

\begin{abstract}
In this work we examine a system consisting of a confined one-dimensional arrangement of atoms that we describe by using the 2-dimensional ${\mathbb C}P^{N-1}$ model, restricted to an interval and at finite temperature. We develop a method to 
obtain the bulk and boundary parts of the one-loop effective action as a function of the effective mass of the fluctuations. The formalism has the advantage of allowing for a systematic analysis of a large class of boundary conditions and to model the (adiabatic) response of the ground state to changes in the boundary conditions. In the case of periodic boundary conditions, we find that inhomogeneous phases are disfavored for intervals of large size. 
Away from periodic boundary conditions, our numerical results show that the ground state has a generic crystal-like structure that can be modulated by variations of the boundary conditions. The results presented here could be relevant for experimental implementations of nonlinear sigma models and could be tested by lattice numerical simulations.
\end{abstract}

\maketitle

\newpage

\textit{Introduction.} A remarkable framework linking high energy with atomic physics has been growing thanks to exceptional developments in cold atoms \cite{ultracold}. The connections are often sparked by the possibility of implementing quantum field theoretical models used in high energy physics with atomic lattices at low temperature and cleverly designed atomic traps \cite{Zohar:2016}. 
An example recently discussed is that of alkaline-earth atoms with SU$(N)$ spins arranged on a two-dimensional bipartite square lattice (with the transverse direction much smaller than the longitudinal one) \cite{Laflamme}. One of the many interesting features of this setup is that its continuum low-energy limit is a known toy model for QCD, the ${\mathbb C}P^{N-1}$ model, a $(1+1)$-dimensional field theory consisting of $N$ complex scalar fields subject to a constraint \cite{Polyakov:1975rr, Polyakov:1975yp,Bardeen:1976zh,Brezin:1976qa}. Despite the apparent simplicity, the ${\mathbb C}P^{N-1}$ model has a very complex vacuum structure featuring asymptotic freedom, dynamical mass generation and confinement, making it an attractive tool to simulate complex physics at a reduced theoretical and computational cost.
While it should come as no surprise that the ${\mathbb C}P^{N-1}$ model emerges from a SU$(N)$ spin model (see, for example, Refs.~\cite{Beard,Kataoka}), a realization with cold atoms provides a concrete possibility of this correspondence, testable in experiments. 

A one-dimensional realization of a \textit{finite} string of atoms, like that of Ref.~\cite{Laflamme}, provides territory to both the condensed matter and the high energy physicist to explore fundamental aspects of the phase structure of cold atomic lattices on one side, and the properties of the underlying continuum field theory (in the present case the ${\mathbb C}P^{N-1}$ model confined to an interval) on the other. In fact, the latter has been at the center of recent discussions revolving around the nature of the ground state, the role of boundary conditions and the possibility that the ground state may develop spatial variations \cite{Monin:2015xwa,Milekhin:2016fai,Gorsky:2013rpa,Bolognesi:2016zjp,Betti:2017zcm,Bolognesi:2018njt,Pikalov:2017lrb,Nitta:2018yen,Flachi:2017xat,Bolognesi:2019rwq,Pavshinkin:2019bed,Chernodub:2018rv,Gorsky:2018lnd}.
These issues become interesting, certainly more amusing, in the original formulation of the model: if we recall that ${\mathbb C}P^{N-1}$ is a $(1+1)$-dimensional model defined on the infinite line, then a well known result \cite{CHMW} prohibits the existence of (and therefore the transition to) a massless phase. While the exis\-ten\-ce of inhomogeneous phases is to be generically expected when the model is restricted to an interval, it is less obvious whether there is any natural choice of boundary behavior allowing for spatially constant phases \cite{Milekhin:2016fai}. The simplest choice of periodic boundary conditions leads, unsurprisingly, to a spatially constant massive ground state for intervals of large size, with a transition to a massless phase predicted, in the leading large-$N$ approximation, when the interval shrinks to a small enough size. The similarity between finite size and temperature effects, that in the case of periodic boundary conditions manifests itself as an exact modular symmetry $\left\{\mbox{size} \leftrightarrow \mbox{inverse temperature}\right\}$, does indeed suggest that a transition might take place, possibly as an artifact of the large-$N$ approximation, occurring below a certain critical size \cite{Monin:2015xwa}. On the basis of the no-go theorem of 
Ref.~\cite{CHMW}, such a massless phase may be discarded \textit{a posteriori} as non physical, leaving, as the only alternative, a spatially constant massive phase (see Refs.~\cite{Hong:1994}). An intriguing possibility has been discussed recently in Ref.~\cite{Gorsky:2018lnd}, where the surprising circumstance of a spatially inhomogeneous ground state has been entertained, still within the original version of the model (in the absence of boundaries or external fields). A different viewpoint has been indicated in Ref.~\cite{Bolognesi:2019rwq}, where the appearance of a massless phase is precluded by a logarithmic behavior emerging after complete integration over the (constant) background fields. 

An additional piece to the story comes from the large-$N$ volume independence of the ${\mathbb C}P^{N-1}$ model with the temporal direction compactified to a circle \cite{Sulejmanpasic:2016llc}. With twisted boundary conditions imposed on the fields and special choice of twists selected, it is only the ground state that contributes to the partition function at any compact radii, leading to observables independent of the radius of compactification and eliminating the possibility of a transition occurring for circles of small enough size. The impossibility of a transition induced by finite size effects then follows inevitably from the large-$N$ volume independence, however it is not evident what happens in general, beyond the specific choice of boundary conditions. 

These conclusions leave the nature of the ground state (and the role of boundary conditions) obfuscated and our goal here is to re-examine the ${\mathbb C}P^{N-1}$ model in the presence of boundaries and with finite temperature effects included. The setup we have in mind 
consists of a string of atoms sandwiched between two impurities or constrained by optical traps that we assume can, in principle, be tuned locally near the endpoints of the string and act as external forcing for the boundary conditions (see Ref.~\cite{jaksch} for an introduction on how defects can be introduced in in optical lattice systems). Then, we expect that changes in the boundary conditions will induce deformations in the ground state of the system: it is these deformations that we are after. 
Beyond the case of periodic boundary conditions, the inevitable complication one needs to address is to allow for a spatially dependent effective mass. To deal with this, we adopt (and adapt to the present problem) a formulation of the effective action based on zeta function regularization and heat-kernel techniques (see Refs.~\cite{Toms:1992dq,Flachi:2010yz}) from which we obtain the effective action in the form of a derivative expansion for the effective mass. The advantage of this approach is its generality that allows to treat, in a unified way, a large class of boundary conditions, of which periodic or Dirichlet are special cases.


\textit{Basics.}  
The classical action of the ${\mathbb C}P^{N-1}$model is 
\bea
\EuScript S = \int dx dt \left[
\left| \partial_{\mu} n_i \right|^2 + \lambda \left(r - \left| n_i \right|^{2}\right)
\right]
\label{cpaction}
\eea 
where $n_i$ with $i=1, \cdots, N$ are $N$ complex scalar fields. Imposing gauge invariance on the model allows one to set any non-dynamical gauge field to zero.
The quantity $\lambda$, that we call the mass gap function, is a Lagrange multiplier that enforces the condition $\left| n_i \right|^{2} = r $. Finally, $r$ defines the coupling constant $g$ by the expression $r=4\pi/g^2$. 
Here, we will stick to the large-$N$ expansion and choose to separate the fields $n_i$ into a classical background plus a quantum fluctuation:
\bea
n_1 &=& \sigma,~~
 n_i = \delta\varphi_i, \mbox{~~~~~~for $i=2, \cdots, N$}. 
\label{bgf}
\eea
We restrict the background to the interval $x \in [0,~\ell]$, thus effectively forcing the fields to obey at the end-points certain boundary conditions that we leave, for the moment, unspecified. We assume $\sigma$, the {\it `Higgs'} field, and $\lambda$ to be real and time-independent functions of the spatial coordinate only. 
Whenever spatially constant configurations for $\lambda$ and $\sigma$ are admissible, the solution $\lambda=0$ \& $\sigma \neq 0$ defines a massless (or Higgs) phase, while  $\lambda \neq 0$ \& $\sigma = 0$ defines a massive (or confining) phase. 

Performing the following coordinate transformation,
\bea
x \rightarrow \tilde x &=& x/\ell,~~~
t \rightarrow \tilde t = t/ \ell
\label{rescaling}
\eea
we can rescale the length of the interval to unity.
Setting $\hlambda =\ell^2 \lambda$, performing a Wick rotation 
and choosing the background fields as in (\ref{bgf}), we arrive at the following expression for the one-loop effective action (at large-$N$),
\bea
\EuScript  S^{E}_{\mbox{\tiny{eff}}} &=& 
\int_0^{\beta/\ell} d\tilde \tau \int_{0}^1 d\tilde x 
\left\{ 
\left({\partial \sigma \over \partial \tilde x}\right)^2 + \hlambda \left( \left| \sigma \right|^{2} - r 
\right)\right\}
\nonumber\\
&&
+{(N-1)} \mbox{Tr} \ln \left(-\Box + \hlambda\right),
\eea
with $\Box = -{\partial^2 / \partial \tilde \tau^2} - {\partial^2 / \partial \tilde x^2}$. Notice we are assuming both $\sigma$ and $\hlambda$ to be functions of $\tilde x$ only. 
At finite temperature
\bea\label{gugli1}
\mbox{Tr} \ln \left(- \Box + \hlambda\right)
\rightarrow
\sum_{n=-\infty}^{\infty}
\ln \det \left({\partial^2 \over \partial \tilde x^2} +  \ell^2 \omega_n^2 + \hlambda \right)
\eea
where 
$\omega_n = {2\pi n / \beta}$, $T=2\pi /\beta$. 
The right-hand-side of (\ref{gugli1}) is a  
divergent quantity, which can be regularized using zeta function regularization (see e.g. Refs.~\cite{Gilkey,Avramidi,ParkerToms,Kirsten,elizalde94,byt}). 
If we write
\bea
\left({\partial^2 \over \partial \tilde x^2} + \hlambda \right) f_k = p_k^2 f_k
\label{sturm}
\eea
we can define the following zeta function, suitable for the regularization of  
(\ref{gugli1}) 
\bea
\zeta (s) =\sum_{k=0}^{\infty}
\sum_{n=-\infty}^{\infty}
\left(p_k^2 + \ell^2 \omega_n^2\right)^{-s},
\label{zeta}
\eea
which is 
well-defined for $\textrm{Re}(s)>1/2$ and can be analytically continued to the semi-plane $\textrm{Re}(s)\leq 1/2$ as a meromorphic function
with isolated simple poles \cite{Gilkey,Avramidi,ParkerToms,Kirsten,elizalde94,byt}. After performing the analytic continuation, the effective action is written as:
\bea
\EuScript  S^E_{\mbox{\tiny{eff}}} &=& 
{\beta} \int_{0}^\ell dx 
\left\{ 
\left({\partial \sigma \over \partial x}\right)^2 + \lambda \left( \left| \sigma \right|^{2} - r \right) \right\} 
-(N-1)\zeta'(0).
\label{effact}
\eea


\textit{Derivative expansion of the effective action.} 
The expression (\ref{effact}) is only formal, as the boundary conditions have not been specified, and a renormalization scheme must be used to correct the diverging terms coming from the summation in (\ref{gugli1}). Our approach consists in recasting the zeta function (\ref{zeta}) in the following form (see \cite{Flachi:2010yz}):
\bea
\zeta(s) &=& 
{\vartheta\over \sqrt{4\pi}}{1\over  \Gamma(s)} \int_0^\infty {dt\over t^{3/2-s}} \EuScript K(t) \left(1+2 \sum_{n=1}^\infty 
e^{-{\beta^2 n^2 \over 4 \ell^2 t}}
\right)
\label{zetafac}
\eea
where $\vartheta =\beta/\ell$ and $\EuScript K(t) = \sum_{k} e^{-t p_k^2}$.
The function $\EuScript K(t)$ is the integrated heat-kernel of the second order differential operator defined in (\ref{sturm}). For the moment, information on the boundary conditions is encoded in the eigenvalues $p_k$. A derivative expansion of the effective action can be read off from the small-$t$ expansion of the heat-kernel (density) that we can express in the form:
\bea
\hat{\EuScript K}(\tilde x, t) = {1\over \sqrt{4 \pi t}} \sum_{k=0}^\infty c_{k/2}\left(\tilde x\right) t^{k/2}.
\label{Kt}
\eea
The quantity $\hat{\EuScript K}(\tilde x, t)$ is related to $\EuScript K(t)$ by spatial integration. The coefficients $c_{k/2}$ depend on the function $\hlambda$ and its derivatives, but not on $t$. The coefficients $c_{k/2}\left(\tilde x\right)$ have the same `dimension' (here the word `dimension' refers to the number of powers of $\hlambda$ and of $\tilde x$), as $\left[ \hlambda \right]^{k/2}$ and comprise a volume plus a boundary contribution for coefficients of integer order, and a pure boundary term for coefficients of half-integer order \cite{Gilkey,Avramidi,ParkerToms,Kirsten,elizalde94,byt}. The first few coefficients are: $c_{0}\left(\tilde x\right) = 1,\; c_{\frac{1}{2}}\left(\tilde x\right) = b^{(\frac{1}{2})}_{0},\; c_{1}\left(\tilde x\right)=-\hlambda + b^{(1)}_{0}$, where the contributions $b_{n}^{(\frac{k}{2})}$ correspond to the boundary parts and have the form $b_{n}^{(\frac{k}{2})} = u^{(\frac{k}{2})}_{n/2} \delta(\tilde x) + v^{(\frac{k}{2})}_{n/2} \delta(\tilde x - 1)$ with $u^{(\frac{k}{2})}_{n/2}$ and $v^{(\frac{k}{2})}_{n/2}$ being numbers solely dependent on the type of boundary conditions. Assuming that boundary conditions are imposed symmetrically, $u^{(\frac{k}{2})}_{n/2} = v^{(\frac{k}{2})}_{n/2}$. Although expressions of such coefficients become quickly cumbersome, partial simplification can be achieved in specific cases. Moreover, we notice that there are terms in the heat kernel coefficients that are independent of $\hlambda$ and can, therefore, be dropped since they contribute neither to the effective equation nor to the boundary conditions. 
For example, $b^{(\frac{1}{2})}_{0}$, $b^{(1)}_{0}$, $b^{(\frac{3}{2})}_{0}$, etc. are all irrelevant to our problem. In the case of periodic boundary conditions all the coefficients $b_{n}^{(k/2)}$ are identically zero. The volume contribution to the coefficients can be found in a number of references (see, for example, \cite{Bastianelli:2008vh,Fucci:2013kma}), while the boundary contributions can be only in part retrieved from the literature (for instance, in Ref.~\cite{Bastianelli:2008vh} for the case of Dirichlet and Neumann boundary conditions and  Ref.~\cite{Fucci:2013kma} for more general boundary conditions). {The heat-kernel coefficients $c_{2}$, $c_{5/2}$, $c_{3}$ have been computed here for the case of separated boundary conditions for the first time.}

A lengthy calculation leads to the following expression for the derivative expansion of the effective action truncated up to order 6, after rescaling everything back to the original variables\footnote{{Notice that we have rescaled everything back to the original variables after a shift $x \to x - \ell/2$; without performing this coordinate shift, the limit $x \to \infty$ would return the case of the semi-infinite interval $\left[0,\infty \right\}$}}:
\bea
\EuScript  S^E_{\mbox{\tiny{eff}}} &=& 
{\beta} \int_{-\ell/2}^{\ell/2} dx \left\{\left({d \sigma \over dx}\right)^2 + \lambda \left( \left| \sigma \right|^{2} - r \right) + \EuScript L\right\},
\label{afaccia1}
\eea
where we have defined $\EuScript L =  (N-1)\;\Big( \EuScript L_0  + \EuScript L_1 + \EuScript L_2\Big)$ with
\bea
\EuScript L_0  &=& {a_0} + {a_2} \lambda + a_4 \lambda^2+ a_6 \lambda^3 +\dots \nonumber \\
\EuScript L_1  &=& {\nu_1} \left({\partial \lambda \over \partial x}\right)^2
+{\nu_2} {\partial^2 \lambda \over \partial x^2}
+{\nu_3} {\partial^4 \lambda \over \partial x^4}
 +\dots
\nonumber \\
\EuScript L_2  &=& 
\xi_0 + \xi_1 \lambda + \xi_2 \ell^2 \lambda^2  
+{\xi_3} \ell {\partial \lambda \over \partial x}
+{\xi_4} \ell \beta^2 \lambda {\partial \lambda \over \partial x} \nonumber\\
&&+{\xi_5} \ell^2 \beta^2\left({\partial \lambda \over \partial x}\right)^2
+{\xi_6} \ell^2{\partial^2 \lambda \over \partial x^2}
+{\xi_7} \ell^3 {\partial^3 \lambda \over \partial x^3}
+\dots. \nonumber
\eea
where the coefficients $\xi_n$ are functions of the dimensionless ratio $\vartheta= \beta/\ell$ that defines our expansion parameter (relevant explicit expressions are reported in appendix). The above terms isolate the powers of the mass gap function in $\EuScript L_0$ from its derivatives in $\EuScript L_1$ and from the boundary terms in $\EuScript L_2$. The functional form of the above expansion can be obtained directly by dimensional considerations. {We should remark that the mixed-derivative expansion is in-built in the heat-kernel expansion and exclude the possibility of ground state rapidly varying in space. While this is a limitation of the present approach, we expect any rapidly varying solution to have higher free energy.}

In our scheme the divergences are encoded in the coefficients $a_0$, $\xi_0$ and $a_2$ and respectively renormalize the vacuum energy in the bulk, on the boundary, and the coupling constant $r$. Also, $a_0$, $\xi_0$ do not depend on $\lambda$ and therefore change neither the effective equations nor the boundary conditions. Explicit expressions of the volume and boundary coefficients up to order 6 are reported in the appendix. Renormalizing to zero the vacuum energy in the bulk and on the boundary (results will not depend on this choice), and noticing that the terms proportional to $\nu_2$ and $\nu_3$ are total derivatives, which disappear after partial integration (since we are assuming boundary conditions to be imposed symmetrically), we obtain the energy density functional: 
$\EuScript  \Omega = \EuScript  \Omega_{\mbox{\tiny vol}}+ \EuScript  \Omega_{\mbox{\tiny bnd}}$, 
where
\bea
\EuScript  \Omega_{\mbox{\tiny vol}} &=& 
\left( \left| \sigma \right|^{2} - r_{ren} +{(N-1)\over 2\pi} \log\left({\beta\over\ell}\right)\right)\lambda
\nonumber \\ 
&& 
-{\zeta(3)\over 16 \pi^3}\beta^{2}\lambda^2   +   {\zeta(5)\over 256 \pi^5}\beta^4 
\left(
\lambda^3
+{1\over 2}  \left({d \lambda\over dx}\right)^2
\right),~~~~~~~~
\label{effacc}
\eea
corresponds to the volume contribution, with $- r_{ren} = - r +(N-1)\left[{\gamma_E \over 4 \pi} - {1\over 4 \pi} \log \left({\varepsilon\over 8\pi^2}\right)\right]$ is the renormalized coupling.
The quantity $\EuScript  \Omega_{\mbox{\tiny{eff}}}^{\mbox{\tiny bnd}}$ corresponds to the boundary action 
\bea
\EuScript  \Omega_{\mbox{\tiny{eff}}}^{\mbox{\tiny bnd}} &=&  
\Big\{\hat \xi_1 \lambda + \hat \xi_2 \ell^2 \lambda^2  
+{\hat \xi_3} \ell {\partial \lambda \over \partial x}
+{\hat \xi_4} \ell \beta^2 \lambda {\partial \lambda \over \partial x} +{\hat \xi_5} \ell^2 \beta^2\left({\partial \lambda \over \partial x}\right)^2 \nonumber\\
&&
+{\hat \xi_6} \ell^2{\partial^2 \lambda \over \partial x^2}
+{\hat \xi_7} \ell^3 {\partial^3 \lambda \over \partial x^3} + \dots
\Big\}\times \Delta_\ell (x),
\label{azionebordo}
\eea
where we have factorized out $\Delta_\ell (x) = \delta\left({x/\ell -1/2}\right) + \delta\left({x/\ell +1/2}\right)$ and the hatted quantities $\hat \xi_i$ correspond to the $\xi_i$ with the factor $\Delta_\ell (x)$ removed. The expansion (\ref{effacc}) allows to easily anticipate, within the validity of our approximation, the character of the ground state and how this depends on the external conditions (in the present case, temperature, interval size, and type of boundary conditions).


\textit{Solutions and boundary conditions.} 
As first step, we vary $\ell$ at fixed temperature ignoring the boundary contribution to the energy density functional and dropping the derivative term in (\ref{effacc}). This situation corresponds to the case of periodic boundary conditions. Then, the thermodynamic potential, including terms up to sixth order, takes the form: $\Psi_{\sf eff} = a_2 \lambda + a_4 \lambda^2 + a_6 \lambda^3$. The coefficient $a_6 =  {\zeta(5)\over 256 \pi^5}\beta^4$ is positive, guaranteeing boundedness from below of the potential. The coefficient $a_4 = - {\zeta(3)\over 32 \pi^3}\beta^2$ is negative signaling an eventual first order phase transition when $a_4 = 2\sqrt{\left|a_2 a_6\right|}$ where $a_2 = \left| \sigma \right|^{2} - r_{ren} +{(N-1) \over 2 \pi} \log\left(\beta / \ell\right)$. Assuming $\sigma =0$, the critical point is located at $\beta_{crit} = \ell \exp\left(-2\pi (r_{ren} + \zeta^2(3)/(8\pi \zeta(5))) /(N-1)\right)$. When $a_4<2\sqrt{\left|a_2 a_6\right|}$ the minima of the potential occurs for $\lambda\neq 0$ leading to a massive constant ground state. When the length scale is decreased below the above critical point, up to $a_4 = \sqrt{\left|3 a_2 a_6\right|}$, then the potential develops two minima, one corresponding to $\lambda \neq 0$ and one, at lower energy, corresponding to $\lambda =0$. While the latter would not be physical as it can be argued by invoking the no-go theorem of Ref.~\cite{CHMW}, the former would be, in principle, admissible. In the case of periodic boundary conditions, Ref.~\cite{Bolognesi:2019rwq} observed that a logarithmic dependence of the energy as a function of an effective constant radius emerges once full integration over the constant modes in the path integral is carried out. It is this logarithmic behavior that would remove the zero mode and block the transition to a massless phase. 
While we can conclude that for periodic boundary conditions and large values of $\ell$ inhomogeneous ground states are disfavored, in the present approach a mechanism that removes the unwanted massless mode remains to be found. 



\begin{figure}
\begin{center}
\begin{tabular}{cc}
  \includegraphics[scale=0.32]{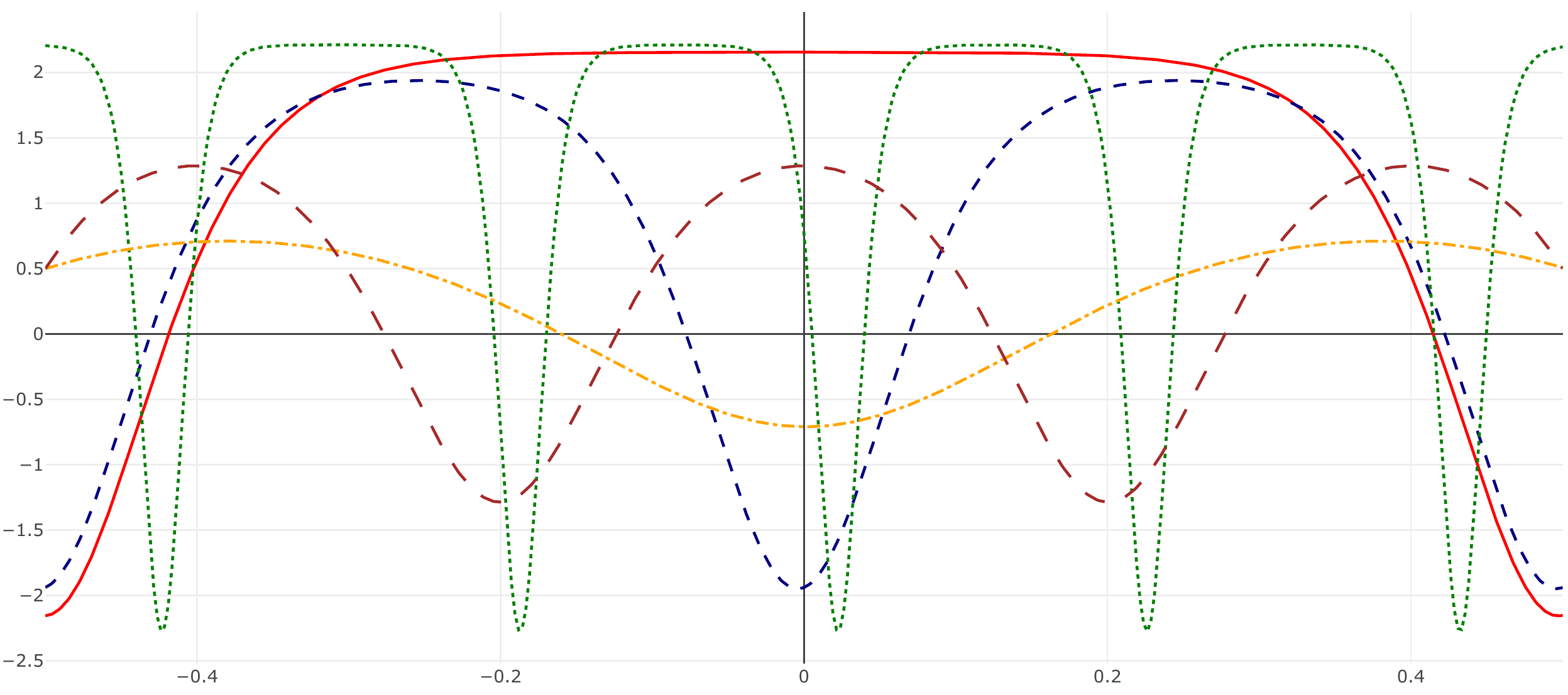}
\put(-120,-10){$x/\ell$}
\put(-240,50){\rotatebox{90}{$\breve{\lambda}$}}
\end{tabular}
\end{center}
\caption{A sample of the numerical solutions. Parameters have been chosen as follows: $r=1$ for all curves (this sets our units) and: $\left\{\beta,\ell, \lambda_-,\lambda'_-\right\} =\left\{1.27,1,-141.5,50.01066\right\}$ (red) continuous curve, $\left\{1.27,1.1,-142,250.987\right\}$ (blue) dashed curve, $\left\{1.2,4.2,115.941,-10.430\right\}$ (green) dotted curve, $\left\{10,10,0.,0.0865\right\}$ (yellow) dot-dashed curve, $\left\{1.27,1,0,1380.1\right\}$  (brown) long-dashed curve.}
\label{figura_soluzioni}
\end{figure}

Here, the focus is on the model confined to an interval with general boundary conditions. In this case, we can no longer ignore the boundary contributions (since, rather than the potential, one needs to \textit{extremize} the effective action; see, for instance, Ref.~\cite{ParkerToms,Toms:1982}). To address this problem, we shall restrict to the small $\sigma$ regime; to lowest order we ignore its contribution to the effective equations and proceed by solving numerically for $\lambda$. Although these equations can be, with some effort, integrated exactly with the solutions expressed in terms of combinations of Jacobi elliptic functions, here we prefer to proceed numerically as it allows for an easier way to browse through the parameter space and the boundary conditions. Within our framework, the boundary conditions, so far left unspecified, can be determined from the boundary action in terms of the coefficients $b_i^{(j)}$. In other words, specifying the boundary conditions fixes the coefficients $b_i^{(k)}$ that, in turn, fixes the boundary action and the behavior of the $\lambda$ at the endpoints (if one ignores higher derivative terms in the boundary action, the boundary conditions for $\lambda$ are in general of mixed-Robin form). Rather than focusing on specific choices of boundary conditions, our numeric approach allows us to browse through the solutions changing the external parameters and the coefficients defining the boundary conditions. In practice, the numerical solutions are found by reducing the nonlinear second order equation of motion for $\lambda$ into two coupled nonlinear first order equations (this approach is also tested by finding the solution directly via solving the second order problem). The (two parameter family of) boundary conditions are implemented by fixing the value $\lambda_-\equiv\lambda(-\ell/2)$ at the left boundary and by changing the derivative $\lambda'_-\equiv\lambda'(-\ell/2)$ at the same point until the solution attains the same value at the right boundary (the numerical tolerance is set to $\lambda_- - \lambda_+ = 10^{-3}$). Once the numerical solutions are found, the free energy is evaluated numerically. The code returns the lowest energy configuration, compatible with the boundary conditions. The boundary conditions can be changed and the procedure repeated. While numerically we can only explore a finite part of the parameter space (and therefore we cannot exclude that other solutions exist outside the considered parameter region), we generically find the ground state characterized by a number of oscillations that we can modulate by changing the parameters and the values of the boundary conditions. It is tempting to conjecture this crystal-like structure of the ground state to be generic. Fig.~\ref{figura_soluzioni} shows a sample of the solutions (we plot the normalized value $\breve{\lambda} \equiv {\lambda - |\lambda| \over |\lambda|}$, with $|\lambda|$ representing the amplitude of the solution for a various values of the parameters.

\textit{Conclusions.} Starting with the ${\mathbb C}P^{N-1}$ model confined to an interval of size $\ell$, we have obtained
a derivative expansion of the energy density functional at one-loop in the presence of general, two-parameter family of boundary conditions. This setup can be used to model a one-dimensional bounded arrangement of atoms and changes in the boundary conditions could be induced by the presence of impurities or by locally modulated optical traps. The general results for the bulk and boundary parts of energy density functional are given, the latter in terms of numerical coefficients that encode the nature of the boundary conditions and can be varied according to the physical set-up. 

This approach has the advantage of allowing for an immediate inspection of some of the properties of the ground state. 

We have considered a general 2-parameter family of boundary conditions, that we assumed to be imposed symmetrically. Physically, what we have in mind is to force external changes in the boundary conditions which, in turn, would cause deformations in the ground state. The response to such changes is investigated by finding numerically the solutions for the mass gap function $\lambda$. These solutions have been constructed for a range of parameters and boundary conditions. Our results show that the ground state is inhomogeneous with a typical crystal-like structure that depends on the imposed boundary conditions and the value of the parameters. The situation is reminiscent of what happens in the Gross-Neveu model at finite density and tempting to conjecture that a similar phase structure can be induced here by modulations in the boundary conditions (see \cite{Flachi:2013bc}). 

The results presented here could be relevant in implementations of the ${\mathbb C}P^{N-1}$ model with alkaline-earth atoms in optical lattices of the kind discussed in Ref.~\cite{Laflamme}. Experimental verifications and lattice simulations could provide a way to verify the existence of spatial modulations in the ground state. Both the approach and the results can be contrasted with other results derived, in the absence of boundaries, for the Gross-Neveu or Nambu-Jona Lasinio models (see Refs.~\cite{Nickel:2009ke,Boehmer:2007ea,Flachi:2010yz}) and for the $O(3)$ nonlinear sigma model (see Ref.~\cite{Senechal:1993rc}). 
Especially interesting are the examples discussed in Ref.~\cite{Senechal:1993rc} for the $O(3)$ nonlinear sigma model where a comparison with experimental observations of the Haldane gap in the one-dimensional phase of $CsNiCI_3$ and of $NENP$ (a Heisenberg anti-ferromagnet with isolated $Ni^{2+}$ chains; see Ref.~\cite{Buyers:1992}) has been carried out and for which qualitative agreement has been indicated. In these cases too, with boundaries enforced, the same kind of modulations in the ground state are expected to emerge. 

Numerical lattice simulations of the ${\mathbb C}P^{N-1}$ model have been actively pursued in the past (see, for example, Refs.~\cite{Bruckmann:2014sla,Bruckmann:2015sua,Berg:1981er,Campostrini:1992ar,Alles:2000sc,Flynn:2015uma,Abe:2018loi,Bonanno:2018xtd}). {Extending the lattice approach to the case treated here (see Ref.~\cite{itou} for such an attempt) could be shed light on some of the questions discussed in this paper and in particular reveal the role of corrections to the large-$N$ approximation}.

\begin{acknowledgements}
This work is partially funded by the Ministry of Education, Culture, Sports, Science MEXT-Supported Program for the Strategic Research Foundation at Private Universities `Topological Science' (Grant No.\ S1511006), the Japanese Society for the Promotion of Science Grants-in-Aid for Scientific Research KAKENHI (Grant n. 18K03626 (AF), 16H03984 and 18H01217 (MN), 19K14616 (RY)), and by MEXT-KAKENHI Grant-in-Aid for Scientific Research on Innovative Areas ``Topological Materials Science'' No. 15H05855 (MN). AF is especially grateful to Toshiaki Fujimori for discussions and to him and Etsuko It\=o for sharing their results prior to publication. Thanks are also extended to Matthew Edmonds, Daisuke Inotani, Bjarke Gudnason, Giacomo Marmorini, Keisuke Ohashi and Vincenzo Vitagliano for discussions. 
\end{acknowledgements}

\appendix

\section*{Appendix: Heat-kernel coefficients}
\label{app:hkc}
In this appendix we provide the reader with the coefficients of the small-$t$ asymptotic expansion of the trace of the heat kernel associated with a one-dimensional Laplace operator of the form given in Eq.(6) in the main text, acting on suitable functions over the interval $[0,1]$. The method used to compute these coefficients for general boundary conditions can be found in Ref.~\cite{Fucci:2013kma}. Here we give the coefficients for the case of general separated boundary conditions. These conditions can be written as 
\begin{eqnarray}\label{separated}
A_{1}f_{k}(0)-A_{2}f_{k}'(0)&=&0\;,\nonumber\\
B_{1}f_{k}(1)-B_{2}f_{k}'(1)&=&0\;,
\end{eqnarray}  
where $f(x)$ are functions belonging to the domain of the Laplace operator given in Eq.(6) (see main text) and the coefficients $A_{i}$ and $B_{i}$ satisfy the constraints $(A_{1},A_{2})\neq (0,0)$ and $(B_{1},B_{2})\neq (0,0)$. Following the results detailed in Ref.~\cite{Fucci:2013kma}, it is not difficult to obtain the following
heat-kernel coefficients associated to the Laplace operator in Eq.(6) (see main text) endowed with the boundary conditions (\ref{separated})
\begin{subequations}
\bea
c_{0}\left(\tilde x\right) &=& 1,\;\nonumber\\ 
c_{\frac{1}{2}}\left(\tilde x\right) &=& b^{(\frac{1}{2})}_{0},\;\nonumber\\
c_{1}\left(\tilde x\right)&=&-\hlambda + b^{(1)}_{0},\;\nonumber\\
c_{\frac{3}{2}}\left(\tilde x\right) &=&  b^{({3\over 2})}_{0}+ {\hlambda} b^{({3\over 2})}_{1},\;\nonumber\\
c_{2}\left(\tilde x\right) &=& {1\over 2} \hlambda^2 -{1\over 6} {\partial^2 \hlambda\over \partial \tilde x^2} +
b^{({2})}_{0} + b^{({2})}_{1} {\hlambda} + b^{({2})}_{2} {\partial \hlambda \over \partial \tilde x},\;\nonumber\\
c_{\frac{5}{2}}\left(\tilde x\right) &=&  b^{({5\over 2})}_{0} + b^{({5\over 2})}_{1} {\hlambda} + b^{({5\over 2})}_{2} {\hlambda^2} + b^{({5\over 2})}_{3} {\partial \hlambda \over \partial \tilde x}+ b^{({5\over 2})}_{4} {\partial^2 \hlambda \over \partial \tilde x^2},\;\nonumber\\
c_{3}\left(\tilde x\right) &=& -{1\over 6} \hlambda^3 + {1\over 12} \left({\partial \hlambda \over \partial \tilde x}\right)^2+ {1\over 6} \hlambda {\partial^2 \hlambda \over \partial \tilde x^2} -  {1\over 60} {\partial^4 \hlambda \over \partial \tilde x^4}\nonumber\\
&& + b^{({3})}_{0} +b^{({3})}_{1} {\hlambda} +b^{({3})}_{2} {\hlambda}^2 
+ b^{({3})}_{3} {\partial \hlambda \over \partial \tilde x}
+ b^{({3})}_{4} {\partial^2 \hlambda \over \partial \tilde x^2}
\nonumber\\
&&+ b^{({3})}_{5} \hlambda {\partial \hlambda \over \partial \tilde x}+ b^{({3})}_{6}  {\partial^3 \hlambda \over \partial \tilde x^3}.\;\nonumber
\eea
\end{subequations}
While the volume part of the heat kernel coefficients is universal, the boundary part depends explicitly on the boundary conditions imposed on the system.
Since we want to provide formulas for the boundary part of the heat kernel coefficients which are valid for any type 
of separated boundary condition (\ref{separated}), we need to introduce a number of functions which will allow us to present a unified treatment of all separated boundary conditions. 
To this end we introduce the following projector 
\begin{equation}
\pi_{\tilde x}=\left\{\begin{array}{ll}
1 & \textrm{if}\; \tilde x=0\\
0 & \textrm{if}\; \tilde x\neq 0
\end{array}\right.\;,
\end{equation} 
and the functions $\chi(\tilde x)$ and $S(\tilde x)$, defined on the boundary of the interval $[0,1]$, having the values
\begin{eqnarray}
\chi(0)&=&1-2\pi_{A_{2}}\;,\nonumber\\
\chi(1)&=&1-2\pi_{B_{2}}\;,
\end{eqnarray}
and 
\begin{eqnarray}
S(0)&=&(1-\pi_{A_{2}})\frac{A_{1}}{A_{2}}\;,\nonumber\\
S(1)&=&(1-\pi_{B_{2}})\frac{B_{1}}{B_{2}}\;.
\end{eqnarray}
Lastly, we have $\Pi_{\pm}(\tilde x)$ defined as
\begin{equation}
\Pi_{\pm}(\tilde x)=\frac{1}{2}\left(1+\chi(\tilde x)\right)
\end{equation}
By further introducing the generalized function $\Delta(\tilde x)=\delta(\tilde x)+\delta(\tilde x-1)$,
one can find
\begin{subequations}
\bea
b^{(\frac{1}{2})}_{0}&=&\frac{1}{4}\chi(\tilde x)\Delta(\tilde x)\;,\nonumber\\
b^{(1)}_{0}&=&2S(\tilde x)\Delta(\tilde x)\;,\nonumber\\
b^{(\frac{3}{2})}_{0}&=&\frac{1}{2}S^{2}(\tilde x)\Delta(\tilde x)\;,\quad b^{(\frac{3}{2})}_{1}=b^{(\frac{1}{2})}_{0}\nonumber\\
b^{(2)}_{0}&=&-\frac{4}{3}S^{3}(\tilde x)\Delta(\tilde x)\;,\quad b^{(2)}_{1}=b^{(1)}_{0}\nonumber\\
b^{(2)}_{2}&=&\frac{1}{3}\left[2\Pi_{+}(\tilde x)-\Pi_{-}(\tilde x)\right]\Delta(\tilde x)\;,\nonumber\\
b^{(\frac{5}{2})}_{0}&=&\frac{1}{4}S^{4}(\tilde x)\Delta(\tilde x)\;,\quad b^{(\frac{5}{2})}_{1}=-b^{(\frac{3}{2})}_{0}\;,\quad b^{(\frac{5}{2})}_{2}=\frac{1}{2}b^{(\frac{1}{2})}_{0}\nonumber\\
b^{(\frac{5}{2})}_{3}&=&-\frac{1}{8}b_{0}^{(1)}\;,\quad b^{(\frac{5}{2})}_{4}=-\frac{1}{4}b^{(\frac{1}{2})}_{0}\;,\nonumber\\
b_{0}^{(3)}&=&-\frac{8}{15}S^{5}(\tilde x)\Delta(\tilde x)\;,\quad b_{1}^{(3)}=-b_{0}^{(2)}\;,\quad b_{2}^{(3)}=-\frac{1}{2}b_{0}^{(1)}\;,\nonumber\\
b_{3}^{(3)}&=&\frac{4}{3}b^{(\frac{3}{2})}_{0}\;,\quad b_{4}^{(3)}=\frac{1}{6}b^{(1)}_{0}\;,\quad b_{5}^{(3)}=-b^{(2)}_{2}\;,\nonumber\\
b_{6}^{(3)}&=&\frac{1}{30}\left[7\Pi_{+}(\tilde x)+12\Pi_{-}(\tilde x)\right]\Delta(\tilde x)\;.\nonumber
\eea

\end{subequations}
The coefficients appearing in the derivative expansion of the effective action (see expressions after Eq.~(12) in the main text) are:
\begin{subequations}
\bea
a_0 &=& -{1\over 8\pi {\varepsilon}} - {\pi \over 3} \vartheta^{-2} \nonumber\\
a_1 &=& {\gamma_E \over 4 \pi} + {1\over 2 \pi}\log\left({\vartheta\over 4 \pi}\right)^2 - {1\over 4 \pi} \log \left(2\varepsilon\right)
\nonumber\\
a_2 &=& -{\zeta(3)\over 32 \pi^3}\beta^{2},\nonumber\\
a_3 &=& {\zeta(5)\over 256\pi^5} \beta^{4}\nonumber\\
\nu_1 &=& {\zeta(5)\over 512\pi^5} \beta^{4}\nonumber\\
\nu_2 &=& {\zeta(3)\over 32\pi^3} \beta^{2}\nonumber\\
\nu_3 &=& {\zeta(5)\over 512\pi^5} \beta^{4}\nonumber
\eea
and
\bea
\xi_1&=& -{b_1^{(3/2)}\vartheta\over 24\sqrt{\pi}} - {\zeta(3)\over 16{\pi^3}}b_1^{(2)}\vartheta^2 -{b_1^{(5/2)}\vartheta^3\over 1440\sqrt{\pi}}
- {3\zeta(5)\over 128{\pi^5}}b_1^{(3)}\vartheta^4\nonumber\\
\xi_2&=& -{1\over 1440\sqrt{\pi}}b_2^{(5/2)}\vartheta^3 - {3\zeta(5)\over 128{\pi^5}}b_2^{(3)}\vartheta^4\nonumber\\
\xi_3 &=&   -{\zeta(3)\over 16{\pi^3}}b_2^{(2)}\vartheta^2 -{1\over 1440\sqrt{\pi}}b_3^{(5/2)}\vartheta^3\nonumber\\
\xi_4 &=&   -{3\zeta(5)\over 128{\pi^5}}b_5^{(3)}\vartheta^2 \nonumber\\
\xi_5 &=&   -{3\zeta(5)\over 128{\pi^5}}b_3^{(3)}\vartheta^2 \nonumber\\
\xi_6 &=&   -{1\over 1440\sqrt{\pi}}b_4^{(5/2)}\vartheta^3 - {3\zeta(5)\over 128{\pi^5}}b_4^{(3)}\vartheta^4 \nonumber\\
\xi_7 &=&   -{3\zeta(5)\over 128{\pi^5}}b_6^{(3)}\vartheta^4 \nonumber
\eea
\end{subequations}

\end{document}